\documentclass[10pt]{iopart}
\usepackage{iopams}
\usepackage{graphicx}
\usepackage{color}
\newcommand{\ket}[1]{|#1\rangle}

\newcommand{\be}{\begin{equation}}
\newcommand{\ee}{\end{equation}}
\newcommand{\bea}{\begin{eqnarray}}
\newcommand{\eea}{\end{eqnarray}}

\newcommand{\fig}[1]{Fig.~\ref{#1}}

\newcommand{\ve}{\varepsilon}

\newcommand{\s}{\sigma}
\newcommand{\G}{\Gamma}
\newcommand{\up}{\uparrow}
\newcommand{\down}{\downarrow}

\begin{document}

\title[Current cross-correlations in quantum dot based Cooper pair splitters]
{Current cross-correlations in double quantum dot based Cooper pair splitters with ferromagnetic leads}

\author{Kacper Wrze\'sniewski, Piotr Trocha and Ireneusz Weymann}

\address{Faculty of Physics,
Adam Mickiewicz University,
61-614 Pozna\'n, Poland}
\ead{wrzesniewski@amu.edu.pl}
\vspace{10pt}
\begin{indented}
\item[]March 2017
\end{indented}

\begin{abstract}
We investigate the current cross-correlations
in a double quantum dot based Cooper pair splitter coupled to
one superconducting and two ferromagnetic electrodes.
The analysis is performed by assuming a weak
coupling between the double dot and ferromagnetic leads,
while the coupling to the superconductor is arbitrary.
Employing the perturbative real-time diagrammatic technique,
we study the Andreev transport properties of the device,
focusing on the Andreev current cross-correlations,
for various parameters of the model,
both in the linear and nonlinear response regimes.
Depending on parameters and transport regime,
we find both positive and negative current cross-correlations.
Enhancement of the former type of cross-correlations indicates transport regimes,
in which the device works with high Cooper pair splitting efficiency,
contrary to the latter type of correlations, which imply negative influence on the splitting.
The processes and mechanisms leading to both types of
current cross-correlations are thoroughly examined and discussed,
giving a detailed insight into the Andreev transport properties of the considered device.
\end{abstract}


%
\vspace{2pc}

\section{Introduction}

Multiterminal quantum dots coupled to superconducting and normal
metal leads have recently been under extensive experimental and theoretical research
\cite{DLoss2000, DLoss2001, feng03, cao, pengZhang, FranceschiNatNano2010, csonka, siqueiraPRB10, konigPRB10_2, rodero11, BraggioSSC, multijunction_nanotube, Droste2015, IW_KPW_PRB15, frolov2016, GramichAppl, hwang2016, domanski2016}.
One of very promising applications of such hybrid systems is the possibility
to realize a Cooper pair splitter (CPS) \cite{russoPRL05, hofstetterSC, herrmann10SC,
hofstetterPRL11, SchindelePRL12, FulopPRB14, TanPRL15, csonka2015arxiv,
yeyati16, Gramich2016, Hussein2016_1, Hussein2016_2}.
In Cooper pair splitting devices the electrons forming a Cooper pair
tunnel from superconductor into two separate normal leads, while the tunability of
quantum dots embedded in the arms of a CPS enables
controlling of the splitting process
\cite{hofstetterSC, herrmann10SC,hofstetterPRL11}.
Cooper pair splitting is most efficient when the applied bias voltage
is smaller than the superconducting energy gap $\Delta$,
such that transport occurs only through Andreev reflection processes
\cite{andreev}.
In fact, the subgap transport properties are determined by the
Andreev bound states (ABS) induced by the
superconducting proximity effect inside the superconducting energy gap.
Such bound states, also referred to as Yu-Shiba-Rusinov states
\cite{Yu-Shiba-Rusinov},
have recently been investigated by performing the bias spectroscopy experiments
\cite{leePRL12,leeNatNano14,schindelePRB14,kumarPRB14}.

The transport properties of hybrid quantum dot systems
in the subgap regime with weak coupling to normal leads have already been a subject of extensive
theoretical investigations \cite{konig09, konigPRB10, bocian, weymannPRB14, wojcikPRB14, trochaPRB14, trochaPRB15, Michalek2016}.
The current flowing through the system inside the superconducting energy gap is dominated by
Andreev processes: crossed Andreev reflection (CAR) and direct Andreev reflection (DAR).
The former processes take place when split electrons reach two separate normal leads,
while the latter ones happen when both electrons enter the same lead.
For the realization of an efficient CPS, it is therefore desirable to
have dominant CAR processes in the system.
Quantifying the relative magnitude of both processes is however not straightforward.
It can be done in quantum dot based CPS's with ferromagnetic contacts,
in which by varying the magnetic configuration of ferromagnetic leads
between the parallel and antiparallel alignment, one can
infer the knowledge about the amount of CAR processes relative to DAR ones
\cite{konig09, weymannPRB14, trochaPRB15, beckmannPRL04}.
In an ideal case of half-metallic leads,
for bias voltages smaller than the dots' charging energy,
the Andreev transport is only due to CAR processes,
and the current is maximized in the antiparallel configuration of leads' magnetic moments,
while in the case of parallel configuration the Andreev current vanishes.
All these indicate that there is a strong motivation
to explore the rich physics of such systems.

In this paper we extend the existing studies
\cite{trochaPRB15}
by analyzing the cross-correlations of the currents flowing
due to Andreev reflection in the CPS based on double quantum dots.
In our model, we consider a double quantum dot system
coupled to an $s$-wave superconductor,
with both dots attached to separate ferromagnetic leads
forming the arms of CPS, see \fig{Fig:scheme}.
We focus on the zero-frequency cross-correlations between
the currents flowing through the left and right ferromagnetic contacts,
defined as \cite{blanter_buttiker}
\be \label{Eq:IS}
S_{LR} = \int_\infty^{-\infty} \!\! dt \langle \delta I_L(t)
                \delta I_R(0) + \delta I_R(0)\delta I_L(t) \rangle,
\ee
where $\delta I_\alpha(t) = \hat I_\alpha(t) - \langle \hat I_\alpha \rangle$
and $\hat I_\alpha$ is the current operator for tunneling between the dot and lead $\alpha$.

The current-current cross-correlations can give a significant insight
into the processes taking place in charge and spin transport \cite{blanter_buttiker},
and have already been successfully measured in various experiments
\cite{SukhorukovNat, mcclure07, Das12}.
In general, an enhancement of positive cross-correlations,
which are strongly present in systems with superconducting electrodes
\cite{bignon04, Melin08, Dong09, Freyn10, ChevallierPRB11, rechPRB12, Nazarov_cross15},
results from interactions supporting currents in both junctions.
In CPS systems it can be associated with high Cooper pair splitting efficiency.
Such positive cross-correlations can be suppressed by interactions, which mutually block the currents,
and/or by tunneling processes that occur in opposite directions.
On the other hand, negative sign of cross-correlations
indicates transport regimes strongly dominated by aforementioned
processes contributing to the total current with different signs.

The present study of spin-dependent transport through double quantum dot system
in the subgap regime is accomplished by means of the real-time diagrammatic technique
\cite{schoeler,thielmann}.
We consider the first-order expansion with respect to the coupling to ferromagnetic leads,
while the coupling to superconductor is arbitrary.
In the CAR transport regime we find strong positive current cross-correlations in a wide range of parameters.
We show that finite hopping between the dots significantly decreases positive cross-correlations
and can result in negative correlations in certain transport regimes.
Moreover, when both CAR and DAR processes are allowed,
both positive and negative current cross-correlations
are present, depending on the transport regime.
We show that the sign of cross-correlations
depends greatly on the energy of Andreev bound states relevant for transport,
which can be tuned by shifting the dots' level positions.
In addition, we analyze the influence of spin polarization of ferromagnets,
showing that tuning the magnetic properties of electrodes is
an important mean to control the splitting efficiency.

We also note that an important application of multiterminal hybrid systems
implementing beam splitter architecture is to perform Bell tests.
There are nevertheless some considerable issues
with predicting entanglement in a straightforward way
from current cross-correlations, mainly due to the quasiparticle emission
at finite temperatures in detectors \cite{Hannes2008}.
However, there exist theoretical proposals of feasible experimental realizations,
with systems involving ferromagnetic contacts \cite{klobusPRB14} or
spin filtering due to the spin-orbit interaction in carbon nanotubes \cite{Braunecker2013}.
The results presented in this work may thus help in tuning the device
into optimal transport regime for such a test experiment,
and can be especially relevant for devices with ferromagnetic junctions.

The paper is organized in the following way. Section II contains description of the model
and method used for calculations, as well as short description of quantities of interest.
Section III is dedicated to the discussion of numerical results.
First, the pure crossed Andreev reflection regime (Sec. III.A)
is analyzed, where we consider the cases of finite level detuning and interdot hopping.
Then, in Sec. III.B we discuss results for the full parameter space
where both CAR and DAR processes are present.
The work is concluded in Sec. IV.


\section{Model and method}

\begin{figure}[t]
\includegraphics[width=1\columnwidth]{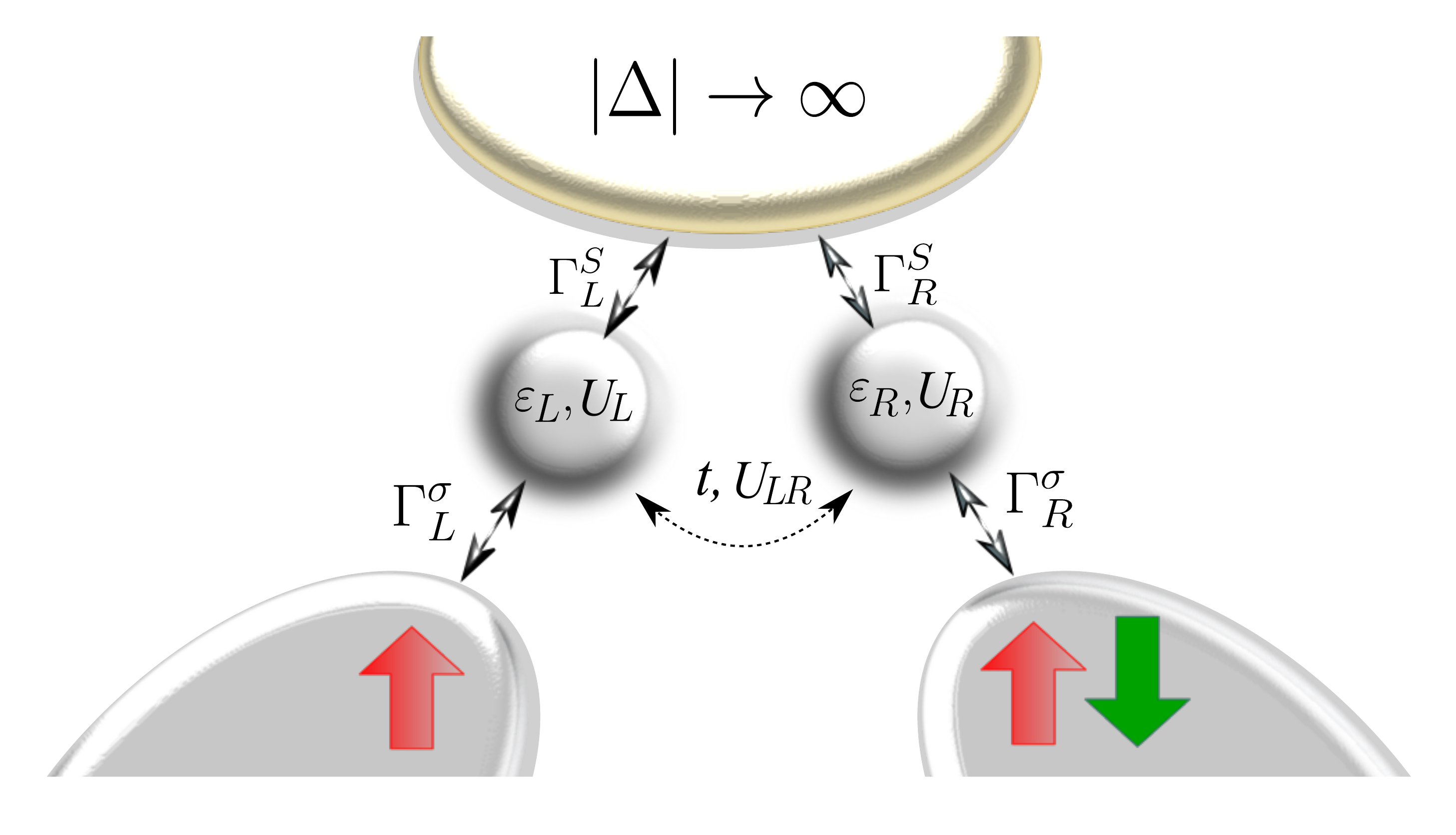}
  \caption{ \label{Fig:scheme}
  Schematic view of the considered system.
  It consists of a double quantum dot coupled to
  two ferromagnetic leads and one superconducting electrode.
  The spin-dependent coupling between the left (right) dot and
  the corresponding ferromagnet is denoted by $\G_{L}^\s$ ($\G_{R}^\s$),
  while the coupling to the superconductor is described by $\G^S_L$ and $\G^S_R$
  for the left and right dot, respectively.
  Each dot is described by on-site energy $\ve_i$ and Coulomb correlation $U_i$,
  whereas capacitive coupling between the two dots is denoted by $U_{LR}$ and
  $t$ is the hopping between the dots.
  It is assumed that the magnetizations of the leads can form
  either parallel or antiparallel configuration, as indicated.
  }
\end{figure}

The schematic of the considered double quantum dot based Cooper pair splitter
with ferromagnetic contacts is shown in Fig.~\ref{Fig:scheme}.
In the central region, two single-level quantum dots are located,
each one tunnel-coupled to a separate ferromagnetic electrode
and both coupled to an s-wave superconductor (top area).
The coupling strength between the $\alpha$-th dot and $\alpha$-th lead
is denoted by $\G_{\alpha}^\s$, with $\alpha=L,R$, for the left and right dot and lead.
On the other hand, the coupling to superconductor
is denoted by $\G_{\alpha}^S$ for the left ($\alpha=L$) and right ($\alpha=L$) dot.
It is assumed that the device can be in two different magnetic configurations:
the parallel $(P)$ and antiparallel $(AP)$ one, as indicated in Fig.~\ref{Fig:scheme}.

\subsection{Effective Hamiltonian}
\label{Sec:H}

The total Hamiltonian of the system consists of four terms
\begin{equation}\label{Eq:1}
  H= H_{DQD}  + H_{FM} + H_S + H_T.
\end{equation}
The first one describes the isolated double quantum dot and is given by
\begin{eqnarray}\label{Eq:3}
  H_{DQD}=\sum_{\alpha=L,R}\Big(\sum_{\sigma} \varepsilon_\alpha d_{\alpha \sigma}^{\dagger}d_{\alpha \sigma}
  +U_\alpha n_{\alpha \uparrow}n_{\alpha \downarrow}\Big) \nonumber \\
   + \sum_{\sigma,\sigma'}\limits U_{LR} n_{L\sigma}n_{R\sigma'}
   +t \sum_{\sigma}(d^{\dagger}_{L\sigma}d_{R\sigma}+d^{\dagger}_{R\sigma}d_{L\sigma}),
\end{eqnarray}
where $d_{\alpha\sigma}^{\dagger}$ $(d_{\alpha \sigma})$ is the creation
(annihilation) operator of an electron in dot $\alpha$
with spin $\sigma$ and energy $\ve_\alpha$.
The on-site Coulomb interaction is denoted by $U_\alpha$,
with $n_{\alpha \sigma} = d_{\alpha\sigma}^{\dagger}d_{\alpha\sigma}$
being the respective occupation number operator.
The interdot Coulomb interaction is described by $U_{LR}$
and $t$ is the hopping between the two dots.

The second term of the Hamiltonian describes the ferromagnetic leads
in the noninteracting quasiparticle approximation,
$H_{FM} = \sum_{\alpha=L,R} \sum_{\textbf{k}\sigma} \varepsilon_{\alpha\textbf{k} \sigma}
c_{\alpha \textbf{k}\sigma}^{\dagger}c_{\alpha \textbf{k}\sigma}$,
where $c_{\alpha \textbf{k}\sigma}^{\dagger}$ is the creation operator
for an electron with spin $\sigma$, wave number $\textbf{k}$ and
energy $\varepsilon_{\alpha \textbf{k}\sigma}$ in the lead $\alpha$.
The superconductor is modeled by the mean-field BCS Hamiltonian
\begin{equation}\label{Eq:2}
H_{S}=\sum_{{\mathbf k}\sigma}  \varepsilon_{S {\mathbf k}\sigma}
     c^\dag_{S {\mathbf k}\sigma}c_{S {\mathbf k}\sigma}
     \!+\!
     \Delta \sum_{{\mathbf k}}\limits\left( c_{S {\mathbf
k}\downarrow}c_{{S -\mathbf k}\uparrow} + {\rm h.c.}\right),
\end{equation}
with $c^\dag_{S {\mathbf k}\sigma}$ being the respective
creation operator for an electron with momentum ${\mathbf{k}}$,
spin $\sigma$ and energy $\varepsilon_{S {\mathbf k}\sigma}$.
The superconducting order parameter is denoted by $\Delta$
and assumed to be real and positive.

The tunneling processes between the double dot
and the leads are described by the last term of the Hamiltonian (\ref{Eq:1}),
which is given by
\begin{equation}\label{Eq:4}
H_T=\!\!\!\! \sum_{\alpha =L,R} \sum_{\mathbf{k}\sigma}
    \!\big(V_{\mathbf{k}\sigma}^\alpha c^\dag_{\alpha\mathbf{k}\sigma}d_{\alpha \sigma}+
    V_{\alpha \mathbf{k}\sigma}^S c^\dag_{S \mathbf{k}\sigma}d_{\alpha \sigma} +
   \rm h.c. \big),
\end{equation}
where $V_{\mathbf{k}\sigma}^\alpha$ are the tunnel matrix elements
between the respective dot and ferromagnetic lead
and $V_{\alpha\mathbf{k}\sigma}^S$ are the elements of tunnel matrix
between the dot $\alpha$ and the superconductor.
For the considered model, we assume that these
matrix elements are momentum and spin independent.
Therefore, the coupling strength between given dot and ferromagnetic lead
can be expressed as, $\Gamma^{\sigma}_\alpha = 2\pi |V^\alpha|^2\rho^{\sigma}_\alpha$,
where $\rho^{\sigma}_\alpha$ is the density of states for spin $\sigma$ in lead $\alpha$.
These couplings can be written as,
$\Gamma^{\sigma}_\alpha = (1 + \sigma p_\alpha)\Gamma_\alpha$,
with $\Gamma_\alpha = (\Gamma^{\uparrow}_\alpha + \Gamma^{\downarrow}_\alpha)/2$
and $p_\alpha$ being the spin polarization of the lead $\alpha$.
On the other hand, the coupling between superconductor
and dot $\alpha$ is given by,
$\Gamma^S_\alpha = 2\pi |V_\alpha^S|^2 \rho^S$,
where $\rho^S$ is the density of states of superconductor in the normal state.

For precise study of the subgap transport properties, we take the limit of
infinite superconducting energy gap $\Delta \rightarrow \infty$.
With this assumption, we are able to focus exclusively on the Andreev reflection processes,
suppressing normal tunneling between superconducting lead and quantum dot system.
When $\Delta$ is the largest energy scale in the problem,
the subsystem consisting of the double quantum dot coupled to superconductor
can be modeled by the following effective Hamiltonian~\cite{rozhkov}
\begin{eqnarray}\label{Eq:5}
  H_{DQD}^{\rm eff}&=&H_{DQD} -\sum_{\alpha=L,R}\frac{\Gamma_\alpha^S}{2}\left(d_{\alpha\uparrow}^{\dagger}d_{\alpha\downarrow}^{\dagger}+ {\rm h.c.}\right)
  \nonumber \\
  &&+\frac{\Gamma_{LR}^S}{2}\left(d_{R\uparrow}^{\dagger}d_{L\downarrow}^{\dagger} + d_{L\up}^{\dagger}d_{R\down}^{\dagger}+ {\rm h.c.}\right).
\end{eqnarray}
The superconducting proximity effect is now accounted for
by induced on-dot pairing correlations described by the last two terms.
The first term, proportional to $\Gamma_\alpha^S$,
describes direct Andreev reflection processes.
Cooper pairs can be also split, when each of the electrons
leaves or enters the superconductor through different dot.
Such crossed Andreev reflection processes
are accounted for by the last term of the effective Hamiltonian,
which is proportional to $\Gamma_{LR}^S=\sqrt{\Gamma_{L}^S\Gamma_{R}^S}$.

Although the coupling between each dot and superconductor can be different,
here we focus on the symmetric case and assume
$\Gamma_L^S = \Gamma_R^S \equiv \Gamma_S$.
Moreover, we also assume symmetric couplings to
ferromagnetic leads, $\Gamma_L=\Gamma_R \equiv \Gamma/2$
and $ p_L=p_R \equiv p$, and symmetric dots, $U_L = U_R \equiv U$.
The system is driven out of equilibrium by applying a bias voltage
between the superconducting and ferromagnetic leads.
The electrochemical potential of superconductor is set to zero, $\mu_S=0$,
while both ferromagnetic leads have the same potential
$\mu_L=\mu_R \equiv \mu = eV$. In such setup, for positive bias voltage, $eV>0$,
the electrons tunnel from the normal leads to superconductor,
while for negative bias, $eV<0$, Cooper pairs are extracted
from superconductor and electrons tunnel to the normal leads,
either in a DAR or in a CAR process.
Because the potentials of the ferromagnets are assumed to be equal,
the net current between them vanishes.

The effective Hamiltonian (\ref{Eq:5}) is not diagonal in the local states of the dots,
$\ket{\chi_L,\chi_R}$, where $\chi_L,\chi_R = 0,\uparrow,\downarrow,d$
for empty, singly occupied with spin-up or spin-down and doubly occupied dot level.
This is because the particle-non-conserving on-dot pairing terms
mix the states with different electron numbers.
Thus, we first diagonalize $H_{DQD}^{\rm eff}$
to find its eigenstates $\ket{\chi}$ and eigenenergies $\ve_\chi$.
In a general case the analytical formulas for them are however too cumbersome to be presented here,
therefore, let us discuss the eigenspectrum for the case when the double occupation of each dot is prohibited
and there is no hopping between the dots.
In this case there are altogether $9$ states relevant for transport,
these are: four single-electron doublet states
$\ket{\sigma,0}$, $|0,\sigma\rangle$,
three triplet states
$\ket{T_0}=(\ket{\!\down,\up}+\ket{\!\up,\down})/\sqrt{2}$ and
$\ket{T_{\sigma}}=\ket{\sigma,\sigma}$,
and (due to the proximity effect the empty state $|0\rangle$ is coupled
with the singlet state $\ket{S}=(\ket{\!\!\down,\up}-\ket{\!\!\up,\down})/\sqrt{2}$)
the two states
\begin{eqnarray}\label{Eq:7}
|\pm\rangle=\frac{1}{\sqrt{2}}\left(\sqrt{1\mp\frac{\delta}
{2\varepsilon_A}}\; |0,0\rangle\mp\sqrt{1\pm\frac{\delta}{2\varepsilon_A}} \; |S\rangle\right),
\end{eqnarray}
where $\delta= \varepsilon_L+\ve_R+U_{LR}$ is a level
detuning parameter and $2 \varepsilon_A = \sqrt{\delta^2+2\Gamma_{S}^2}$
is the energy difference between states $|+\rangle$ and $|-\rangle$.
The energies of the singly occupied states are simply given by
$E_{10} = \ve_L$ and $E_{01} =\ve_R$, the energy of triplet states is $E_T=\ve_L+\ve_R+U_{LR}$,
while the eigenenergy of states $\ket{\pm}$ equals $E_{\pm}=\delta/2\pm\varepsilon_A$, respectively.
The excitation energies between the singlet and doublet
states yield the Andreev bound state (ABS) energies
\be
E^{\rm ABS}_{\eta\gamma} = \eta \frac{U_{LR}}{2} + \frac{\gamma}{2} \sqrt{\delta^2+2\Gamma_{S}^2}\;,
\ee
with $\eta,\gamma = \pm$.

We would like to emphasize that in our considerations
we focus on the pure Andreev reflection regime, for which $\Delta>U$.
This condition is however not necessarily
satisfied in arbitrary hybrid quantum dot-superconductor systems.
For instance, in carbon nanotube quantum dots, the charging
energy is typically of the order of $2-4$ meV \cite{multijunction_nanotube},
which is higher than the energy gap $\Delta$ for superconductors such as
Pb ($\sim1.3$ meV) or Nb ($\sim1.2$ meV)
\cite{multijunction_nanotube, GramichAppl}.
Nevertheless, the ratio of $U/\Delta$ can be enhanced in
systems with large gap superconductors,
for which $\Delta$ can reach a few meV \cite{HeinrichNaturePhys}.
Moreover, for nanowire quantum dots the charging energy
is lower than for molecular quantum dots,
and typically is of the order of $1$ meV.
Consequently, there are systems in which the Coulomb correlations
are of the order of the superconducting energy gap.
Therefore, while our results for bias voltages up to the charging energies
are clearly important for current experiments,
they also shed light on the Andreev reflection processes at higher voltages,
which may be of importance for future experiments.

\subsection{Method and quantities of interest}

In this paper we focus on the Andreev transport properties
of double quantum dot based Cooper pair splitters
weakly attached to ferromagnetic leads.
We are in particular interested in cross-correlations
of the currents flowing through the left and right ferromagnetic junctions $S_{LR}$.
Moreover, to make the picture complete and be able
to identify different transport regimes,
we also calculate the Andreev current $I_S$
and associated differential conductance $G_{S} = dI_S/dV$.
To find the Andreev current, we first need to determine the
normal current between the double dot and
the left ($I_L$) and right ($I_R$) ferromagnetic lead.
Then, $I_{S}$ can be simply found from the Kirchhoff's law, $I_{S}=I_L+I_R$.

To determine the current and current cross-correlations
we use the real-time diagrammatic technique
\cite{schoeler,thielmann,palaNJP07,governalePRB08,weymannPRB08}.
This technique relies on systematic perturbation expansion of the reduced density matrix
and relevant operators with respect to the tunneling Hamiltonian.
Here, we perform the expansion with respect to the coupling strength $\Gamma$
to ferromagnetic leads, while the coupling to superconducting electrode is arbitrary.
In calculations we take into account the lowest-order terms of expansion,
which describe sequential tunneling processes between the dots and ferromagnetic leads.
Within the real-time diagrammatic technique, the quantities of interest
can be expressed in terms of self-energies, which can be calculated
by collecting contributions from relevant diagrams.
Once the self-energies are determined,
the stationary occupation probability $p^{st}_{\chi}$
of the eigenstate $\ket{\chi}$ of effective DQD Hamiltonian
can be found from \cite{schoeler,thielmann}
\begin{equation}\label{Eq:master}
   \mathbf{W} \mathbf{ p^{st}}=0 ,
\end{equation}
together with the normalization condition.
Here, $\mathbf{ p^{st}}$ is the vector containing probabilities $p^{st}_{\chi}$
and the elements $W_{\chi\chi'}$ of self-energy matrix $\mathbf{W}$
account for transitions between the states $\ket{\chi}$ and $\ket{\chi'}$.
The current flowing through the junction with ferromagnetic lead $\alpha$
can be found from \cite{schoeler,thielmann}
\begin{equation}\label{Eq:8}
  I_{\alpha}=\frac{e}{2\hbar} {\rm Tr} \left\{ {\mathbf W}^{I_\alpha}{\mathbf p}^{st} \right\},
\end{equation}
where the self-energy matrix ${\mathbf W}^{I_\alpha}$
is similar to ${\mathbf W}$, but it takes into account the number
of electrons transferred through a given junction.

In order to determine the cross-correlations of the currents flowing through the left and right junctions,
first, we have to calculate both matrices ${\mathbf W}^{I_L}$ and ${\mathbf W}^{I_R}$.
The current-current correlation function has contributions from two current operators
which appear either in one irreducible block or in two distinct blocks.
For the cross-correlations in the sequential tunneling approximation,
the former auto-correlation term vanishes, which leads to the following expression for $S_{LR}$
\cite{schoeler,thielmann}
\begin{equation}\label{Eq:Slr}
  S_{LR}=\frac{e^2}{\hbar} {\rm Tr} \left\{ \left[
  {\mathbf W}^{I_L}\mathbf{P} {\mathbf W}^{I_R}+
  {\mathbf W}^{I_R}\mathbf{P} {\mathbf W}^{I_L}\right]
  {\mathbf p}^{st} \right\},
\end{equation}
where the propagator ${\mathbf P}$ is determined
from $\mathbf{\tilde{W}P} = \mathbf{p}^{st}\mathbf{e}^T - \mathbf{1}$,
with $\mathbf{\tilde{W}}$ being identical to matrix $\mathbf{W}$ with arbitrary one row
replaced by $(\Gamma, \Gamma,\dots, \Gamma)$ and $\mathbf{e}^T = (1,1,\dots,1)$.

The perturbation expansion with respect to the coupling to ferromagnetic leads
limits us to the weak coupling regime, with $\Gamma$ being
the smallest energy scale.
Therefore, it is important to notice that within this approximation,
it is not possible to grasp higher-order correlations, such as for instance those
leading to the Kondo effect \cite{glazman89,avishaiPRB03,yeyatiPRM03}.

\section{Results and discussion}

In this section we present and discuss
the transport properties of the considered system
for various parameters of the model.
First, we examine the transport characteristics assuming infinite
on-site Coulomb correlations. With this assumption we are able to focus
exclusively on the CAR transport regime.
We analyze the current cross-correlations
both in the absence and presence of detuning
between the dots' levels, as well as
in the presence of hopping between the dots.
Although both finite detuning and hopping
lead to a similar splitting of Andreev bound states \cite{trochaPRB15},
we show that they have
a completely different influence on the current cross-correlations.
We then allow for double occupation of both dots,
and study the transport behavior
in the presence of both DAR and CAR processes.
Before proceeding, let us recall that the
bias voltage in all considered cases is applied in the same way,
with superconducting lead grounded and equal potential $\mu=eV$
applied to ferromagnetic leads.

\subsection{Crossed Andreev reflection regime}

\begin{figure}[t]
\begin{center}
\includegraphics[width=1\columnwidth]{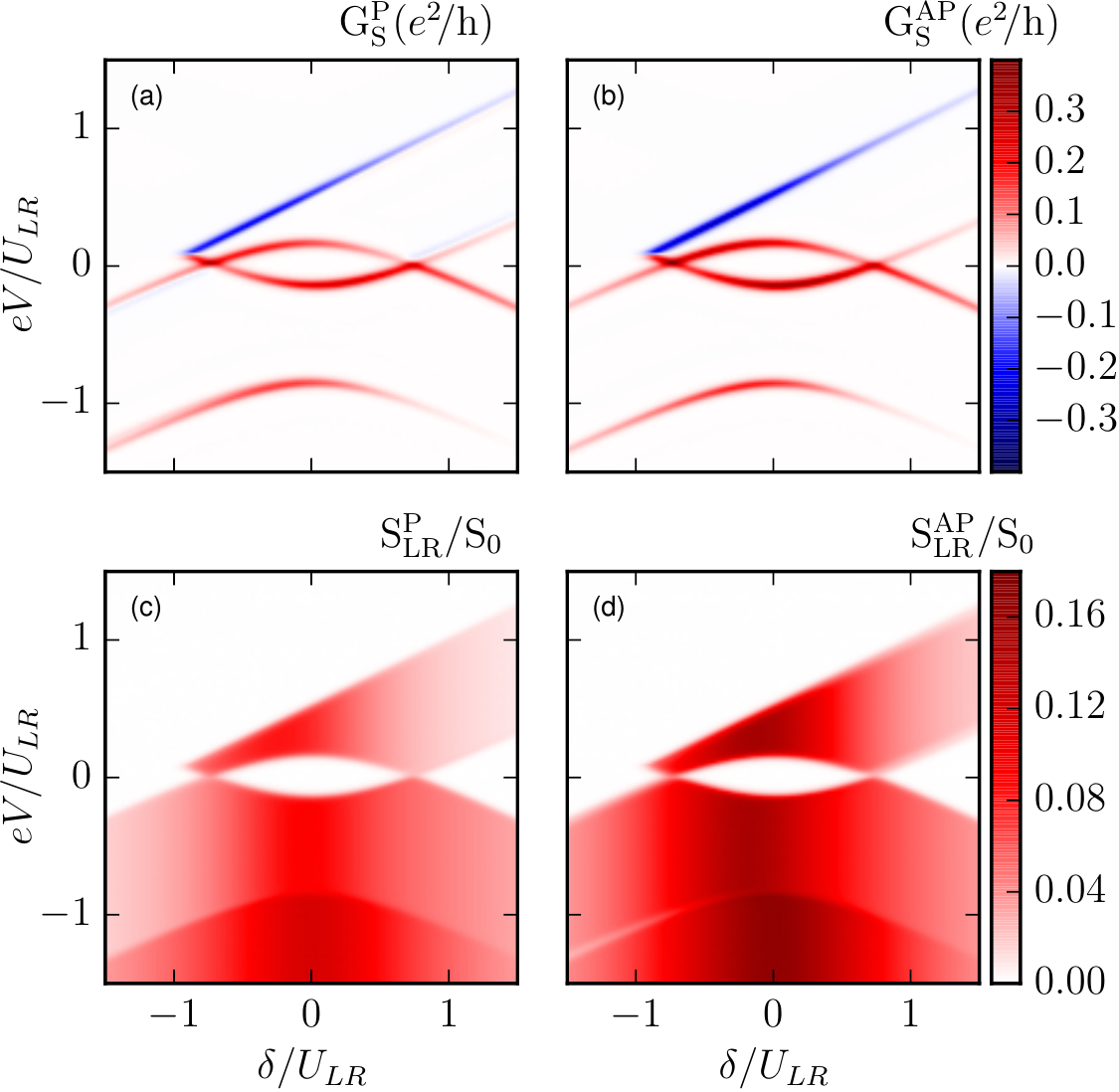}
  \caption{\label{Fig:2}
  (a,b) The differential conductance $G_S = dI_S/dV$
  of the Andreev current
  and (c,d) the current-current cross-correlations
  in the parallel (left column) and antiparallel (right column)
  magnetic configuration as a function of detuning parameter $\delta = 2\ve+U_{LR}$
  and the applied bias voltage $eV$.
  The parameters are: $\G_S=0.5$, $T=0.015$, $\Gamma=0.01$,
  $p=0.5$, $t=0$, $U=\infty$, with $U_{LR}\equiv 1$ the energy unit, and $S_0=e^2\G/\hbar$.
  }
  \end{center}
\end{figure}

When the on-site interaction in the dots
is assumed to be infinite ($U\rightarrow\infty$),
the double occupancy of each dot is prohibited.
There are then $9$ states relevant for transport in this limit,
as discussed in Sec. \ref{Sec:H}.
As a result, the current flows exclusively
due to crossed Andreev reflection processes,
while direct processes are not allowed.
In Fig.~\ref{Fig:2} we show the Andreev differential conductance
and cross-correlations of Andreev currents
as a function of applied bias voltage $eV$
and level detuning parameter $\delta = 2\ve+U_{LR}$,
assuming $\ve_L = \ve_R = \ve$.
The left column presents the density plots for the parallel
magnetic configuration of ferromagnetic leads, while the right
column corresponds to the antiparallel configuration.

Let us first discuss the behavior of the Andreev differential conductance.
The current starts flowing, when the energy of Andreev bound
state lies within the window provided by the bias voltage.
Consequently, at the threshold voltage there is
a peak in differential conductance for both positive and negative bias voltages.
In the low bias voltage regime, for wide range of detuning parameter
we observe the Coulomb blockade where the current is strongly suppressed.
When approaching $|\delta| = \sqrt{U_{LR}^2-2\Gamma_S^2}$ ($\delta/U_{LR} \approx 0.7$ in Fig. \ref{Fig:2}),
the Andreev bound states enter the resonance, resulting in a maximum in the differential conductance.
As can be seen in Fig. \ref{Fig:2}, independently of magnetic configuration, we can identify both
positive and negative differential conductance (NDC) peaks.
When the first bound state becomes active in transport,
the Andreev current starts flowing and there is a positive peak in the differential conductance.
However, the next peak in $G_S$ when increasing the bias voltage
can be either positive or negative, depending on the sign of $eV$.
The NDC forms for positive bias voltage,
starting from $eV/U_{LR}\approx 0$ for detuning parameter $\delta/U_{LR}\approx -1$,
and progressing along the higher values of $eV/U_{LR}$ and $\delta/U_{LR}$
according to $eV \approx (\delta+U_{LR})/2$, see Figs. \ref{Fig:2}(a) and (b).
At this bias voltage, the triplet state occupation becomes strongly enhanced, which leads to the suppression
of the Andreev current, since it is forbidden to tunnel from a triplet state into an $s$-wave superconductor.
This phenomenon is known as the triplet blockade of the Andreev current \cite{konig09, trochaPRB15}.
When the bias voltage is reversed, the electrons from the triplet state
can always tunnel to normal ferromagnetic leads and the triplet blockade is not present.

Contrary to the differential conductance,
the current cross-correlations are non-negative
in the whole range of bias voltage and detuning parameter $\delta$,
irrespective of magnetic configuration, see Figs.~\ref{Fig:2}(c) and (d).
Positive sign of $S_{LR}$ is characteristic of crossed Andreev reflection
processes \cite{Melin08,Freyn10}. When the double occupancy of each dot is forbidden,
extracting/injecting Cooper pairs from/into superconductor
is always associated with the current flowing through both left and right junction,
since this can happen only in a CAR process.
Consequently, the Andreev current flowing through one junction has positive
influence on the Andreev current flowing through the right junction,
and one thus finds $S_{LR} \geq 0$.
We also note that the current cross-correlations are nonzero
only when the sequential current flows through the system.
Therefore, similarly to the current, the cross-correlations
are suppressed in the triplet blockade regime.
Moreover, as can be seen in Fig.~\ref{Fig:2},
enhanced cross-correlations occur
around the particle-hole symmetry point, $\delta=0$,
since then the energy of doubly occupied and empty DQD states
is minimum and the Andreev reflection is maximized.

The main difference between the results in the parallel and antiparallel magnetic configuration
of the device is related with the magnitude of the considered quantities.
Both the differential conductance and the current-current cross-correlations
have higher absolute values in the antiparallel configuration
compared to the parallel one.
This observation indicates that spin-dependent tunneling from and into ferromagnetic leads
has a strong impact on the Andreev transport properties \cite{trochaPRB15}.
To understand enhanced transport in the antiparallel configuration,
it is desirable to take a look at the most important processes active in transport.
Transferring a Cooper pair through the system involves two electrons
of opposite spins. Consequently, in the antiparallel
configuration, there is a fast transport channel, which
involves majority spins of both ferromagnets.
In the parallel configuration, on the other hand,
one of the electrons is always a minority spin,
which results in a smaller overall rate for tunneling processes.
The difference in the currents reveals in
the associated difference in the current cross-correlations,
which are generally enhanced for the antiparallel leads' alignment
compared to the parallel one, $S_{LR}^{AP} > S_{LR}^{P}$.

Assuming the zero temperature limit,
it is possible to derive some approximate analytical formulas
for both $I_S$ and $S_{LR}$.
For $eV<0$ and $\delta \approx0$, the first step in the
current appears between $-1\lesssim eV/U_{LR}\lesssim-0.2$.
In this regime, in the parallel configuration, the current reaches
\begin{equation}\label{Eq:Ip_analytics}
  I^{P}/I_0=\frac{p^2-1}{p^2+3},
\end{equation}
where $I_0=e\Gamma/\hbar$
and the current-current cross-correlations can be approximated by
\begin{equation}\label{Eq:Slr_analytics1_p}
  S_{LR}^{P}=\frac{(-3+p)(-1+p)(1+p)(3+p)(1+3p^2)}{4(3+p^2)^3}\frac{e^2\Gamma}{\hbar}.
\end{equation}
In antiparallel configuration, the current is given by
$I_S^{AP}/I_0=-1/3$, while for the current-current cross-correlations one finds
\begin{equation}\label{Eq:Slr_analytics1_ap}
  S_{LR}^{AP}=\frac{(9+7p^2)}{108(1-p^2)}\frac{e^2\Gamma}{\hbar}.
\end{equation}

For parameters where the current is maximized and exhibits a plateau,
the current-current cross-correlations also retain a finite and constant value.
The current in the parallel configuration
for the plateau at $eV/U_{LR}\lesssim-1$ and $\delta \approx0$
is given by, $I_S^{P}/I_0=p^2-1$,
while the current-current cross-correlations are given by
\begin{equation}\label{Eq:Slr_analytics2_p}
  S_{LR}^{P}=\frac{1}{32}\left[ 4-p^2(1+p^2)^2\right]\frac{e^2\Gamma}{\hbar}.
\end{equation}
On the other hand, in the case of antiparallel configuration
the current is maximized with $I_S^{AP}/I_0=-1$,
and the cross-correlations can be expressed as
\begin{equation}\label{Eq:Slr_analytics2_ap}
  S_{LR}^{AP} = \frac{1}{8-8p^2}\frac{e^2\Gamma}{\hbar}.
\end{equation}

\begin{figure}[t]
\begin{center}
\includegraphics[width=1\columnwidth]{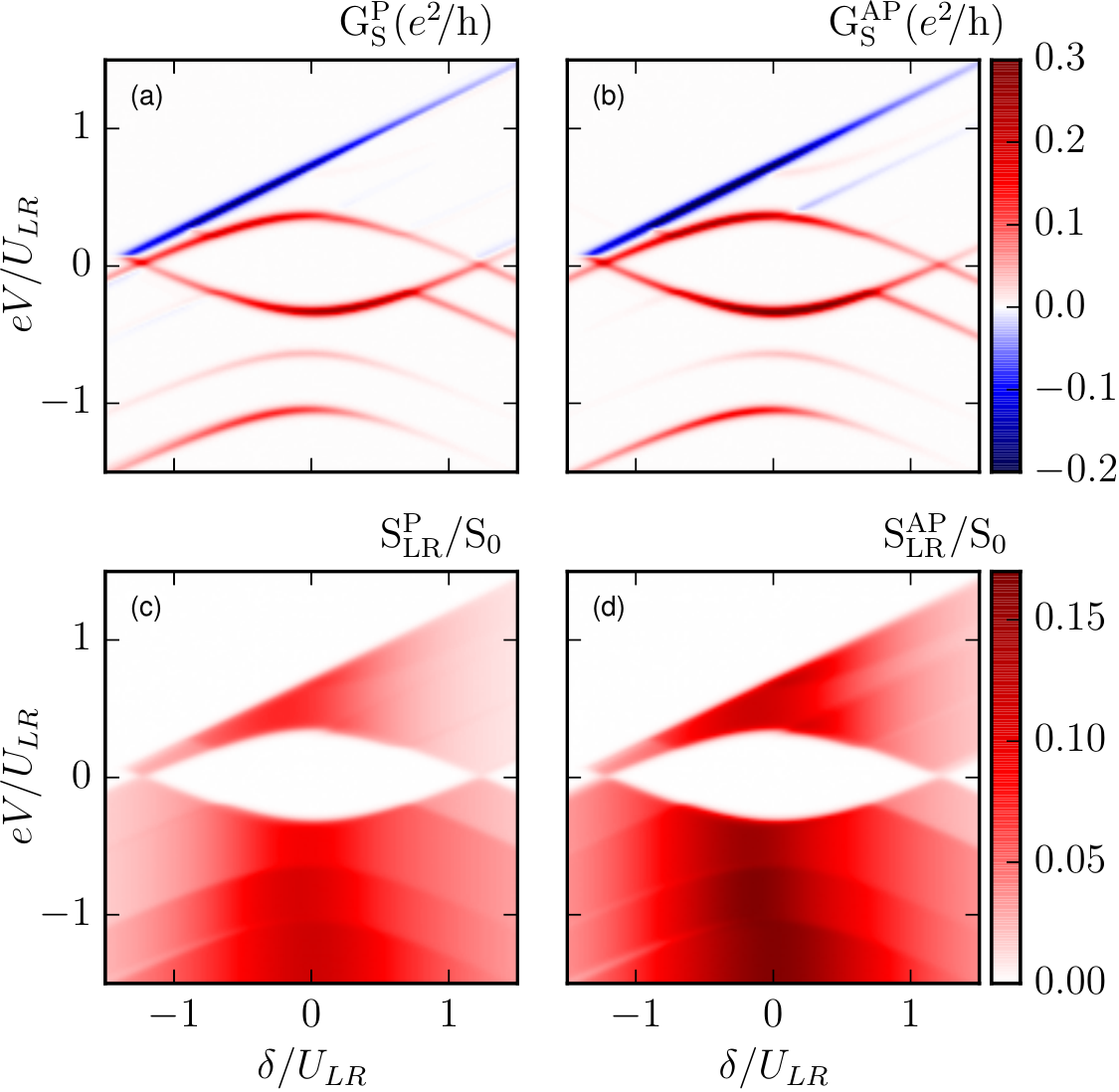}
  \caption{\label{Fig:3}
   The detuning and bias voltage dependence
   of (a,b) the Andreev differential conductance
   and (c,d) the current cross-correlations
   calculated in the case of parallel (left column) and
   antiparallel (right column) magnetic configuration.
   The other parameters are the same as in \fig{Fig:2}
   with finite level detuning $\Delta\varepsilon=\varepsilon_L-\varepsilon_R=0.4 U_{LR}$.}
  \end{center}
\end{figure}

When the double dot levels are detuned by a gate voltage,
$\varepsilon_L\neq \varepsilon_R$,
the Andreev bound states become split \cite{trochaPRB15}.
Figure~\ref{Fig:3} displays the differential conductance and current-current
cross-correlations as a function of $eV$ and $\delta$
in the presence of finite level detuning $\Delta\varepsilon/U_{LR} = 0.4$,
with $\Delta\varepsilon = \varepsilon_L-\varepsilon_R $.
The main difference between the system's transport properties
in the absence and presence of detuning
is related to new peaks visible in the Andreev differential conductance
and corresponding new plateaus in $S_{LR}$.
Except for that, the general behavior
of the current cross-correlations is similar to the case of
$\varepsilon_L = \varepsilon_R$; $S_{LR}$
are positive in the whole range considered in \fig{Fig:3}
and show an enhancement around the particle-hole symmetry point.

\begin{figure}[t]
\begin{center}
\includegraphics[width=1\columnwidth]{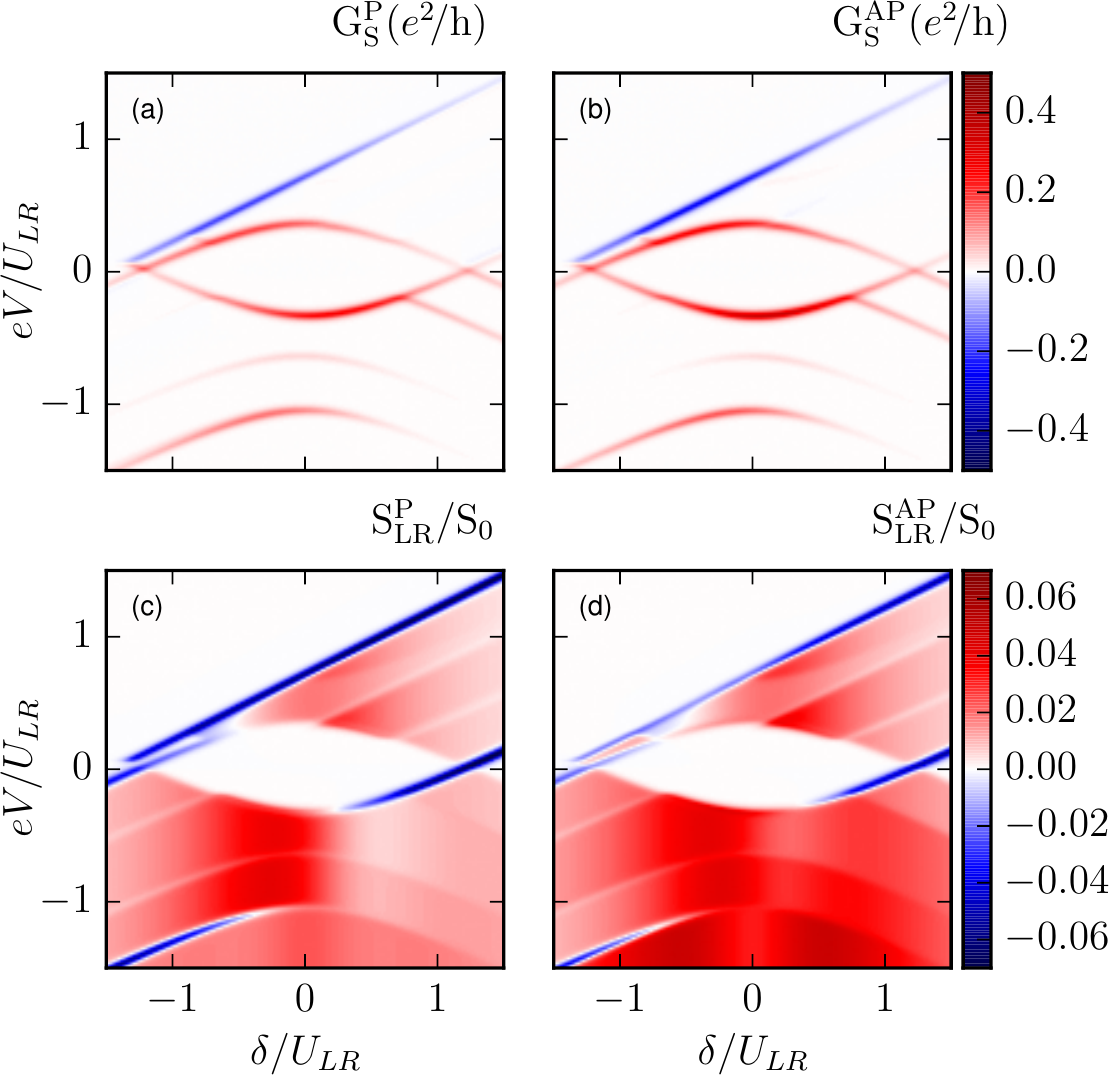}
  \caption{\label{Fig:5}
  (a,b) The Andreev differential conductance $G_S$
  and (c,d) the current cross-correlations $S_{LR}$
  in the case of parallel (left column) and antiparallel (right culomn)
  magnetic configuration calculated
  as a function of detuning $\delta$ and applied bias voltage $eV$.
  The parameters are the same as in \fig{Fig:2} with $t=0.2/U_{LR}$.}
  \end{center}
\end{figure}

The Andreev bound states become also split
when there is a finite hopping between the dots.
In \fig{Fig:5} we show the same transport characteristics
as in \fig{Fig:3}, however now, instead of finite level detuning,
we investigate the influence of the interdot hopping by assuming $t/U_{LR}=0.2$.
A good point to start the analysis is by comparing the Andreev differential conductance in the case
with finite hopping and with finite level detuning, see Figs. \ref{Fig:3} and \ref{Fig:5}.
The results are qualitatively very similar,
i.e. the number of conductance peaks and their position are the same
for $t=\Delta\varepsilon/2$, and the main difference is only related to
slightly different values of the conductance in both cases.

Contrary to the Andreev current, the cross-correlations
display a nontrivial change once $t\neq 0$.
As can be seen in Figs. \ref{Fig:5}(c) and (d),
there is a range of parameters where $S_{LR}$ becomes negative.
This important difference visible in current cross-correlations is
related to the bonding and antibonding states that form due to finite $t$.
The four single-electron states, $\ket{\sigma,0}$, $|0,\sigma\rangle$,
which were present in the case of $t=0$,
now form the following doublets
\begin{eqnarray}\label{Eq:states_finite_hopping}
|\sigma_{\pm}\rangle=\frac{1}{\sqrt{2}}(\ket{\sigma, 0} \pm \ket{0, \sigma}).
\end{eqnarray}
These states are delocalized over both quantum dots and tunneling from/into
these states is now possible through both left and right junctions.
Although this fact has a rather small impact on the steady-state current,
it may result in a more dramatic change in $S_{LR}$,
which is determined by time-dependent correlations between the currents.
As in the case of $t=0$, each split electron entered different dot
and tunneled further to different ferromagnetic lead,
in the case of finite $t$, when one of the two electrons
leaves the double dot through one junction, the other electron,
which is now delocalized, can leave the system through the same
junction. Consequently, the two electrons forming a Cooper pair
are no longer forced to tunnel into separate leads.
More specifically, due to the above-mentioned single-electron states,
non-CAR processes are possible, with electrons sequentially tunneling
through the same junction, one after another.
Such scenario has now a finite probability and
such processes take in fact significant part in transport.
As a result, we observe a drop of positive cross-correlations
in a wide range of transport parameters compared to the previous case
and, more importantly, we find regimes where $S_{LR}$
changes sign and becomes negative.

As can be seen in Figs.~\ref{Fig:5}(c) and (d),
three strong negative cross-correlation lines emerge,
having the same value of a positive slope in the ($\delta$, $eV$)-plane,
aligned along conductance peaks and parallel to each other.
Negative cross-correlations appear for both directions of applied bias voltage.
They result from the tunneling processes in distinct junctions
that contribute to the current with opposite signs.
This sequence of tunneling events is now possible
for certain transport parameters, contrary to the case with vanishing interdot hopping $t$.
Due to the finite $t$,
the one-electron states are no longer localized on single quantum dots,
therefore the effect of the infinite Coulomb interaction
$U\rightarrow\infty$ is not critical for the type of allowed processes,
as it does not block the tunneling of consecutive electrons through the same junction.
Now, from any of single-electron states, there is a possibility to tunnel
through either left or right junction, which was forbidden
in the case of negligible hopping.
For each of the three negative cross-correlation peaks, the energy of the states $|\pm\rangle$
is such that most tunneling processes occur between them
and single-electron states, but in opposite directions, as the relevant states are near
resonance with electrochemical potential of ferromagnetic leads.
The single-electron states are most likely to change into one of the
$|\pm\rangle$ states by an electron tunneling through the normal junction.
However, from the former state, it is energetically more favorable for an electron to tunnel back
to the ferromagnetic lead.
These two processes are dominant for this narrow set of transport
parameters, $eV$ and $\delta$, and when they happen through distinct ferromagnetic junctions
negative cross-correlations emerge.

It is also important to note that negative cross-correlations
may be accompanied by vanishing current $I_S\approx 0$,
which is understandable, as most processes occur in opposite directions,
canceling out the contribution to the average current.
This is however not the case for $eV/U_{LR}\lesssim -1$ and $\delta/U_{LR}\lesssim -0.5$
where negative $S_{LR}$ appears despite finite current flowing through the system.
In this case, there are two one-electron doublets, well separated energetically.
One of them is inside the window provided by the transport voltage
and contributes strongly to the current, however its contribution
to cross-correlations is relatively low.
On the other hand, the second doublet
is around the electrochemical potential of ferromagnetic leads,
and transport processes between these states and Andreev bound states
are negatively correlated, which results in $S_{LR} < 0$.
Final important observation is that negative cross-correlation peak
occurring around the triplet blockade is additionally enhanced
by the fact that from the triplet state, the only possible tunneling
processes are in opposite direction, back into ferromagnetic leads.

\begin{figure}
\begin{center}
\includegraphics[width=1\columnwidth]{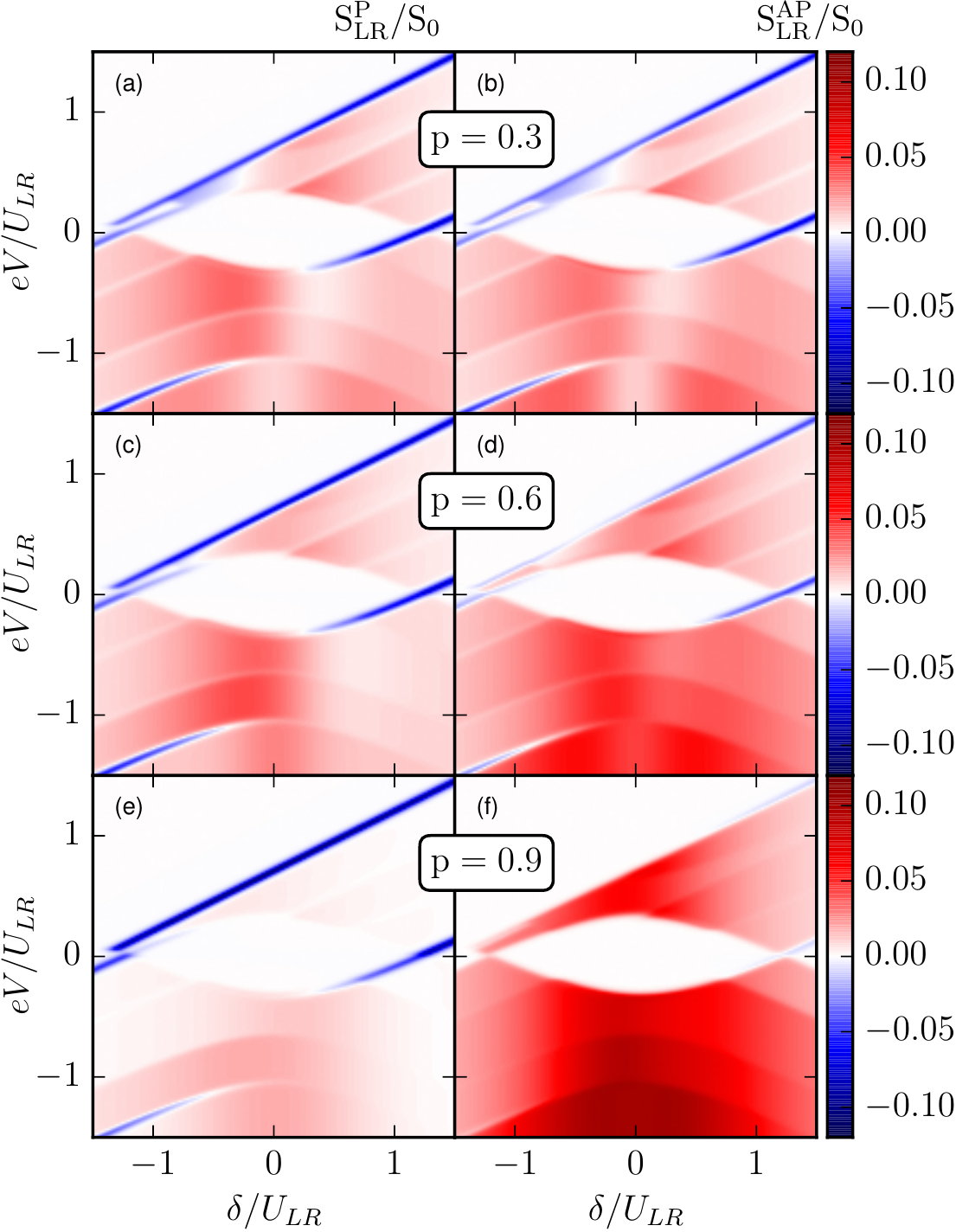}
  \caption{\label{Fig:6}
  The current cross-correlations in the parallel $S^P_{LR}$
  and antiparallel $S^{AP}_{LR}$ magnetic configurations
  as a function of detuning $\delta$ and applied bias voltage $eV$.
  Parameters are the same as in \fig{Fig:2} with $t=0.2/U_{LR}$
  and different values of spin polarization $p$ of ferromagnetic leads,
  as indicated.}
  \end{center}
\end{figure}

Let us now analyze how the magnitude of the electrodes'
spin polarization influences the current cross-correlations.
Figure~\ref{Fig:6} presents the bias voltage and level detuning
dependence of $S_{LR}$ calculated for both
the parallel (left column) and antiparallel (right column) magnetic
configurations of ferromagnetic leads. The spin polarization
$p$ varies from relatively small $p=0.3$ up to $p=0.9$, as indicated.
When $p\lesssim 0.3$, one can see that,
both qualitatively and quantitatively, current cross-correlations
show similar dependence in both configurations,
with slightly higher absolute values in the antiparallel configuration.
However, when the spin polarization increases,
clear differences appear in $S_{LR}^{P}$ and $S_{LR}^{AP}$,
involving not only quantitative, but also qualitative changes.
In particular, one can see for $p=0.9$ [\fig{Fig:6}(e)] that in the parallel configuration
positive cross-correlations are strongly diminished,
while negative ones have a sharper dependence and higher absolute values.
In the antiparallel configuration, on the other hand,
the behavior is quite opposite.
Positive cross-correlations are significantly enhanced,
while the negative ones are almost completely suppressed,
see \fig{Fig:6}(f).

The mechanism explaining the above behavior is strictly associated
with the magnitude of leads' spin polarization and the resulting spin-dependent couplings.
When the magnetizations of the leads form the antiparallel configuration,
the DQD states are strongly coupled to two separate reservoirs
with opposite majority states. Thus, there are high tunneling rates for
both majority spins, taking place in distinct junctions, which results in
strong positive current cross-correlations.
The decrease of spin polarization $p$
lowers the dominant coupling strengths, which in
consequence weakens positive cross-correlations.
Switching the magnetic configuration
to the parallel one,
increases this effect even further.
Now, only electrons with spin aligned with
leads' polarization have high tunneling rates, and the electrons with opposite spin
are weakly coupled to whichever ferromagnetic lead.
Therefore, the escaping processes for one of the electrons forming a Cooper pair
into the ferromagnetic lead is enhanced by relatively high coupling strength,
but the rate for the associated process
of the other electron with opposite spin is relatively low.

The coupling imbalance also strongly influences negative cross-correlations,
but in a reverse way, i.e. the strongest negative cross-correlations occur in the parallel polarized setup,
see \fig{Fig:6}.
In this configuration, for parameters for which
negative cross-correlations occur, the processes giving opposite contributions to the current
most likely take place through distinct leads.
This is completely opposite to the case of strongly-polarized antiparallel configuration
when the negative values of $S_{LR}$ are almost completely suppressed, see \fig{Fig:6}(f).
This is because for consecutive processes in opposite directions,
due to strong polarization of the leads,
an electron of majority spin which tunnels to the DQD,
will most probably tunnel back through the same junction,
as the electron of opposite spin is very weakly coupled.
Such processes do not contribute to cross-correlations
because both tunneling events happen through the same junction.

Once again, we look for analytical approximations
of the current and the current cross-correlations
in the current plateaus in the particle-hole symmetric case.
In both magnetic configurations,
the current for $eV/U_{LR}\lesssim-1$ and $\delta \approx0$
has approximately the same value as
the current in the aforementioned case without hopping.
However, the current cross-correlations are strongly reduced and given by
\begin{equation}\label{Eq:Slr_analytics1_p}
  S_{LR}^{P}=\frac{p^2}{32}(3-2p^2-p^4)\frac{e^2\Gamma}{\hbar}
\end{equation}
for the parallel and
\begin{equation}\label{Eq:Slr_analytics1}
  S_{LR}^{AP}=\frac{p^2}{8}\frac{e^2\Gamma}{\hbar}
\end{equation}
for the antiparallel magnetic configuration.

By a simple comparison with already discussed case of infinite $U$,
we expect that experimental measurement of
cross-correlations can help to estimate how well both quantum dots are separated from
each other. Lack of negative cross-correlations may suggest that dots are not
tunnel coupled. There are already experimental realizations of
Cooper pair splitters using distant graphene quantum dots,
such that no direct hopping between the dots is possible \cite{TanPRL15}.
Examining the current cross-correlations might be then a good indicator, if the device
is well-fabricated and has desired properties.
On the other hand, by detecting negative values of cross-correlations,
one can assume significant finite hopping
influencing transport properties and decreasing splitting efficiency.
However, we would like to note, that this conclusion is only valid
when strong charging energy is present on the dots. In the following
section, we will show that, despite vanishing hopping,
negative current cross-correlations can be measured
when the charging energy is finite and comparable to the applied voltage.

\subsection{Crossed and direct Andreev transport regime}

\begin{figure}
\begin{center}
\includegraphics[width=1\columnwidth]{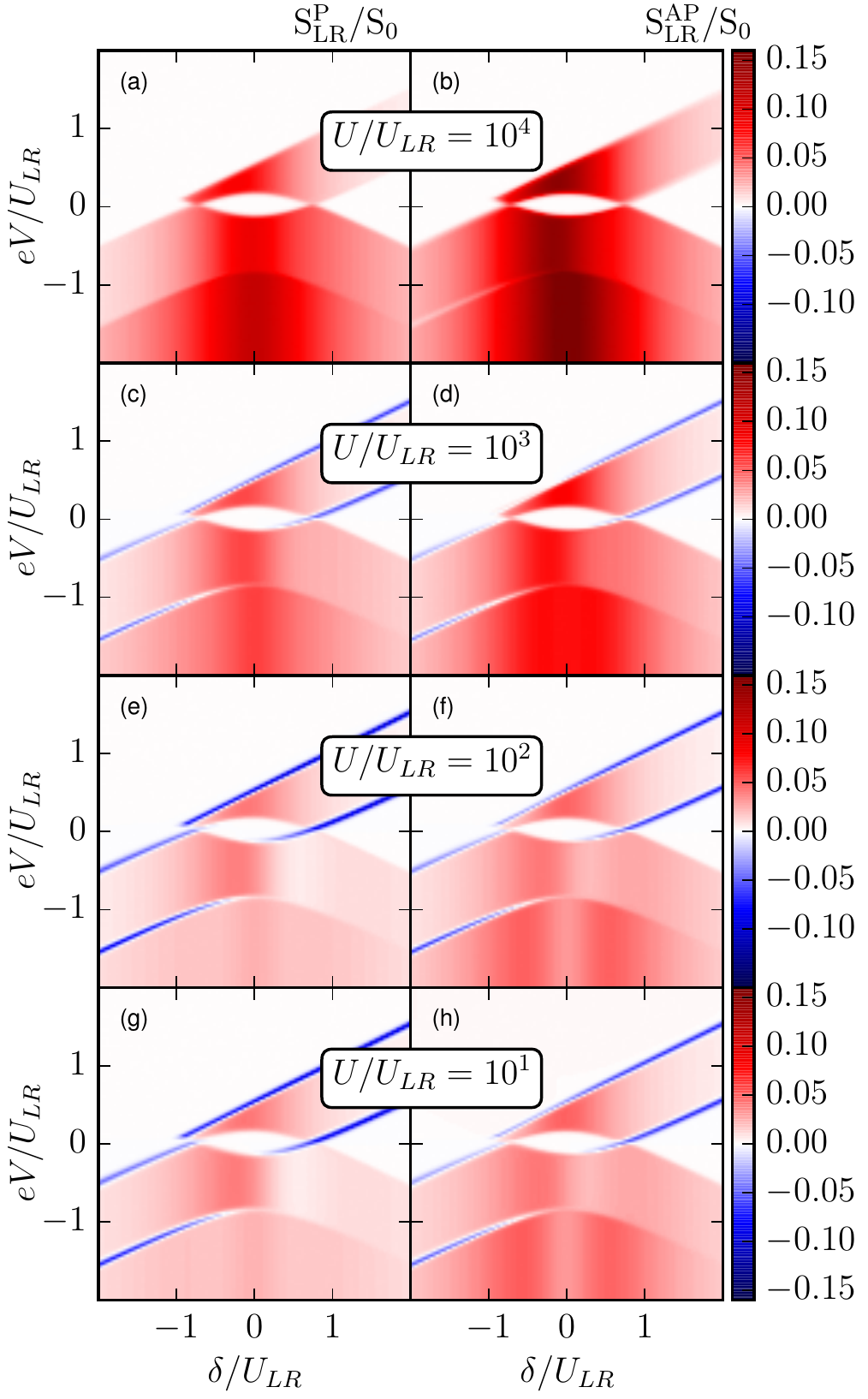}
  \caption{\label{Fig:full_U}
  The current cross-correlations in the parallel $S^P_{LR}$
  and antiparallel $S^{AP}_{LR}$ magnetic configurations
  as a function of detuning $\delta$ and applied bias voltage $eV$
  calculated for different values of on-site Coulomb correlation parameter $U$.
  The other parameters are the same as
  in \fig{Fig:2}.}
  \end{center}
\end{figure}

\begin{figure*}
\includegraphics[width=0.9\textwidth]{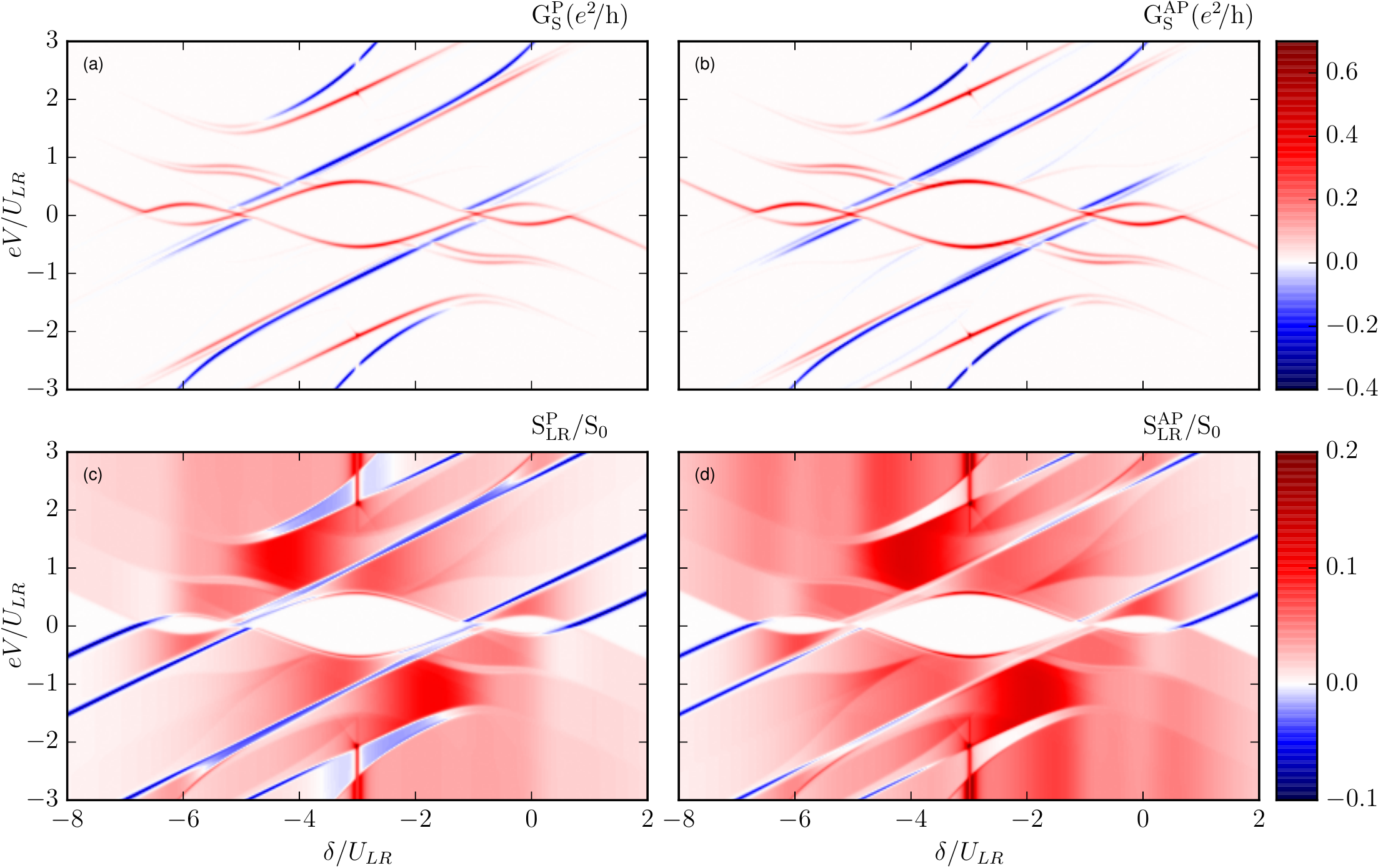}
  \caption{\label{Fig:full}
  The differential conductance $G_S = dI_S/dV$
  of the Andreev current in the parallel ($G^{P}_S$) and antiparallel ($G^{AP}_S$)
  magnetic configurations and current cross-correlations
  for the parallel ($S_{LR}^{P}$) and antiparallel ($S_{LR}^{AP}$) magnetic configurations
  as a function of detuning $\delta = 2\ve+U_{LR}$
  and the applied bias $eV$. The parameters are
  the same as in \fig{Fig:2} with $U/U_{LR} = 2$.}
\end{figure*}

In this section we analyze the effect of finite
on-site Coulomb correlation $U$
and its influence on the current cross-correlations.
Figure~\ref{Fig:full_U} presents $S_{LR}$
in both magnetic configurations as a function of $\delta$ and $eV$
for different values of $U$.
Starting with an extremely high value of on-site correlations,
see the case of $U/U_{LR}=10^4$ in Fig.~\ref{Fig:full_U},
we observe strong positive cross-correlations
in the range where the current is expected to flow
through the system,
suggesting significant presence of CAR processes in transport,
while negative cross-correlations are not present.
This fact supports the statement that one should expect the behavior
of the system with very high $U$ to be essentially the same
as that predicted in the case of $U\rightarrow\infty$, cf. \fig{Fig:2}.
However, one needs to be careful with the above conclusion.
As can be seen in Figs.~\ref{Fig:full_U}(c) and (d),
which present results for $U/U_{LR}=10^3$,
despite $U$ being still the highest, finite energy scale in the problem,
greatly exceeding both bias voltage and thermal energy, negative cross-correlations can appear.
With lowering the strength of Coulomb correlations,
we observe further increase of negative cross-correlations
and decrease of positive ones, see \fig{Fig:full_U}.

To explain this behavior it is important to realize that
the presence of Coulomb correlations strongly influences the states forming
in double quantum dot system. Again,
in the considered transport regime
especially important are the single-electron states.
Let us recall that in the case of $U\rightarrow\infty$ ($t=0$)
the singly occupied states are $\ket{\sigma,0}$, $|0,\sigma\rangle$.
When $U$ is finite this is however not the case
due to on-site pairing correlations in the double dot
generated by the superconducting proximity effect.
For finite value of $U$, those states
have the following general form,
%
$|\sigma_{\pm}\rangle=\alpha\ket{\sigma, 0} \pm \beta\ket{0, \sigma}$,
with $\alpha$ and $\beta$ being parameter-dependent amplitudes.
When $U$ is very large only one of those amplitudes is relevant,
however, for lower values of correlations, both amplitudes are becoming important
and comparable. As a result, single-electron delocalized
states start playing a role similar to the bonding and antibonding
states in the case of finite $t$, giving rise to negative correlations
between the left and right currents.
As can be seen in \fig{Fig:full_U}, lowering the value of $U$
results in an enhancement of negative cross-correlations.
It is important to emphasize that the above mechanism of hybridization
is due to superconducting correlations and Coulomb interaction $U$
is a parameter controlling these effects, as for lower and lower values,
it allows to form more delocalized states in the system.

Finally, let us analyze the behavior of cross-correlations
in the full parameter space assuming $U/U_{LR}=2$.
Figure ~\ref{Fig:full} shows the differential conductance
and current cross-correlations in both magnetic configurations,
for a wide range of applied bias voltage and detuning parameter $\delta$.
All quantities exhibit now symmetry with respect to the particle-hole symmetry point
$\varepsilon_{ph}=-U_{LR}-U/2$ under the replacement $eV\to -eV$ and $\delta \to -\delta$.
For the detuning parameter used in the figure it has the following value,
$\delta_{ph}= -U_{LR} - U = -3U_{LR}$, see Fig.~\ref{Fig:full}.
With lowering the detuning parameter $\delta$, the double dot
becomes consecutively occupied with electrons.
More precisely, the charge in DQD changes when $\delta \approx U_{LR}$,
$\delta \approx -U_{LR}$, $\delta \approx -2U - U_{LR}$
and $\delta \approx -2U - 3U_{LR}$. In between those points,
at low bias voltage the system is in the Coulomb blockade.
The blockade can be lifted when the applied voltage becomes larger than
a certain threshold voltage, such that Andreev bound states
enter the transport window.
Due to more states relevant for transport as compared to the case of
infinite $U$, the differential conductance shows now more peaks
corresponding to additional Andreev bound states \cite{trochaPRB15}.

As far as cross-correlations are concerned,
one can clearly identify the regimes where negative $S_{LR}$
occur. They appear as four lines across
the wide range of bias voltage and detuning parameter,
with a positive slope in the ($\delta, eV$)-plane,
each one parallel to another one,
see \fig{Fig:full}.
Moreover, they are aligned along the strongest conductance peaks
related to new quantum dots' states entering the transport window
with higher occupancy. The mechanism responsible
for the formation of these negative cross-correlation lines
is similar to that discussed in the case of finite hopping $t$.
In present case, despite $t=0$, the states
with odd electron number, which are delocalized over the two dots,
are responsible for negative cross-correlations between the currents.

In addition, an important difference in cross-correlations as compared to the case of large $U$
can be seen around the triplet blockade regimes,
which are near $\delta=0$ for $eV/U_{LR}\gtrsim1$, as well as $\delta/U_{LR} = -6$
for $eV/U_{LR}\lesssim-1$, see Figs. \ref{Fig:full}(c) and (d).
In the case of infinite on-site Coulomb interactions,
the current was fully suppressed due to the triplet blockade, and
so were the current cross-correlations.
In the case of finite $U$, the cross-correlations
have a positive sign in this regime,
indicating a current leakage through the states with doubly occupied dots.

\section{Conclusions}

In this paper we have analyzed the current-current cross-correlations in
a double quantum dot based Cooper pair splitter with ferromagnetic leads.
We focused on the transport regime where the current flows due to Andreev reflection
processes and the calculations were performed by using
the real-time diagrammatic technique.
First, we considered the case of infinite Coulomb on-site
correlations on the dots, when the current is purely due to
crossed Andreev reflection processes. We showed that the cross-correlations
are then positive in the full range of bias voltage and DQD detuning parameter,
and are larger in the case of antiparallel magnetic configuration
compared to the parallel one.
Moreover, we analyzed the effect of finite hopping between the dots
and demonstrated that it can induce negative current cross-correlations
that greatly depend on the degree of spin polarization of the leads
and magnetic configuration of the device.

We also considered the case of finite Coulomb correlations
on the dots and analyzed the transport behavior in the full parameter space.
It turned out that there are regimes of negative
cross-correlations even in the absence of hopping between the dots.
The mechanism responsible for negative correlations
between the currents flowing through both junctions
is generally associated with odd-electron states
that are delocalized over the two quantum dots.
Because positive current cross-correlations
can be related to high Cooper pair splitting efficiency,
the presented results point out the important conditions and parameters
for optimizing the operation of Cooper pair splitters
based on double quantum dots.

\ack

This work was supported by the Polish National Science
Centre from funds awarded through the decision No. DEC-2013/10/E/ST3/00213.


\end{document}